\documentclass[twoside,twocolumn]{article}

\usepackage{blindtext} 

\usepackage{graphicx}
\usepackage{subfigure}

\usepackage[sc]{mathpazo} 
\usepackage[T1]{fontenc} 
\usepackage{microtype} 

\usepackage[english]{babel} 

\usepackage[hmarginratio=1:1,top=32mm,columnsep=20pt]{geometry} 
\usepackage[hang, small,labelfont=bf,up,textfont=it,up]{caption} 
\usepackage{booktabs} 

\usepackage{lettrine} 

\usepackage{enumitem} 
\setlist[itemize]{noitemsep} 

\usepackage{abstract} 

\usepackage{titlesec} 
\renewcommand\thesection{\Roman{section}} 
\renewcommand\thesubsection{\roman{subsection}} 
\titleformat{\section}[block]{\large\scshape\centering}{\thesection.}{1em}{} 
\titleformat{\subsection}[block]{\large}{\thesubsection.}{1em}{} 

\usepackage{fancyhdr} 
\usepackage{titling} 

\usepackage{hyperref} 

\newcommand{\rta}{\rightarrow}

\newcommand{\ep}{\epsilon}

\newcommand{\om}{\omega}
\newcommand{\ra}{\rangle}
\newcommand{\la}{\langle}

\newcommand{\beq}{\begin{equation}}
\newcommand{\eeq}{\end{equation}}

\newcommand{\ball}{\begin{align}}
\newcommand{\eall}{\end{align}}

\newcommand{\beqar}{\begin{eqnarray}}
\newcommand{\eeqar}{\end{eqnarray}}

\newcommand{\ben}{\begin{enumerate}}
\newcommand{\een}{\end{enumerate}}

\pretitle{\begin{center}\huge\bfseries} 
\posttitle{\end{center}} 
\title{An attempt to settle the one-component versus two-component debate in Cuprate high-$T_c$ superconductors} 
\author{%
\textsc{Navinder Singh}\thanks{Cell Phone: +919662680605} \\[1ex] 
\normalsize Physical Research Laboratory, Ahmedabad, India. \\ 
\normalsize \href{mailto:navinder.phy@gmail.com}{navinder.phy@gmail.com} 
}
\date{\today} 
\renewcommand{\maketitlehookd}{%
\begin{abstract}
Related to the electronic structure of cuprate high-Tc superconductors there are two prevailing views: In one view, Pines and collaborators argue that cuprates consist of two electronic subsystems: one with localized spins and the other having itinerant character. Contrary to that, Phil Anderson has argued for one-component or one-band model for cuprates. Thus, a very natural question arrises: What is the actual electronic system in cuprates? Are these two-component or one-component systems? After a careful consideration of both the views, we argue in favor of one-component or Andersonian view. The key lies in the dual character of electrons in narrow d bands, and we put forward a semiclassical model using which we could quantitatively account for oxygen NMR shift data in $La_{2-x}Sr_xCuO_4$ within the single component view.
\end{abstract}
}

\begin{document}

\maketitle


\section{Introduction}

At the foundation of the problem of high-$T_c$ superconductivity in Cuprates is the nature of their electronic structure. More specifically the question is: in which hybrid orbitals the magnetic degrees of freedom reside and where do the mobile carriers reside?  For this fundamental question,  there are two prevailing views in the literature: In one view Pines and collaborators argue\cite{pines}  that cuprates consist of two electronic subsystems or two interpenetrating fluids:  one with localized spins and the other having itinerant or mobile character. The reason for this view is that the NMR shift experimental data can be nicely explained within this two-fluid or two-component picture, as the investigations of Haase, Slichter and collaborators have recently shown\cite{haase,haase2,haase3}.  Microscopically it is explained on the basis of in-complete d-p hybridization of copper $3d$ orbitals and oxygen $2p$ orbitals, and it leads to localized copper spins and mobile $p$ holes (i.e., two components).  With this division magnetic paring mechanism conceptually becomes pleasing as one can imagine mobile carriers being paired up by the magnetic spin fluctuations of the localized copper spins (along the conventional BCS tradition). 

Contrary to that, right from the beginning, Phil Anderson has argued for one-component or one-band model for cuprates as far as low energy physics is concerned\cite{ander1,ander2}. This relevant band is build up from hybridization of copper $d$ orbitals and oxygen $p$ orbitals, more precisely, the anti-bonding $Cu~3d_{x^2-y^2}-O~2p_{\sigma}$ narrow band. This view is based on very careful analysis of the nature of chemical bonding in cuprates. Thus, a very natural question arrises: What is the actual electronic system in cuprates? Are these two-component systems or one-component systems? In this paper we attempt to resolve this issue and put forward a semiclassical model which quantitatively account for the oxygen NMR shift data in LSCO within the single component scenario. The physical picture of our model consists of temporally localized states in Cu $d-$orbitals and a band of mobile states separated from the temporally localized states with a "spin gap". This physical picture at finite doping is some sort of {\it remanent of Mott physics}  at zero doping.

The paper is organized in the following way. We first review Anderson's argument in favor of one component view which is based on very careful analysis of the semi-covalent bonding in Cuprates. We then consider the view-point of Pines and collaborators in favor of two-component view. NMR shift data of Haase and collaborators is reviewed which supports the two-component view. We argue how the debate can be settled and introduce our model which is based on the one-component view and compute oxygen NMR shift from it. On comparing with experimental data we find a reasonably good agreement. We close by summarizing the results.

\section{The correct electronic structure of Cuprates and which electrons are removed on doping?}

In what follows we closely follow Anderson's explanation of the electronic structure of Cuprates\cite{ander1}. Let us take the example of the system $La_{2-x}Sr_xCuO_4$.  The composition of the un-doped system $La_2CuO_4$ is: two layers of $LaO$ and one layer of $CuO_2$. Thus $Cu$ must be in  $Cu^{++}$ state, as $La$ generally has the valence state $La^{+++}$ and oxygen has $O^{--}$. $Cu$ has electronic configuration $[Ar] 3d^{10}4s^1$ and $Cu^{++}$ has to be $[Ar] 3d^9$.  As $d$ orbitals accommodate 10 electrons, thus in one of the $d$ orbitals in $Cu^{++}$ one electron has to remain single or un-paired. There are five d orbitals ($xy,~yz,~zx, ~x^2-y^2, ~3z^2-r^2$). The question is:  in which orbital does this un-paired or lone electron reside?

We will speak in terms of orbitals not hybridized orbitals as $d$ orbitals are comparatively tightly closed in the interior of the atom. They are relatively "sequestered" as compared to $s$ orbitals which hybridize considerably and form fat bands. In the system under consideration, copper atoms are in a cages of octahedrons formed by oxygen atoms (figure 1).  We want to understand how d-orbitals are filled.  With zeroth approximation, crystal fields are of cubic symmetry\footnote{Remember six oxygen atoms, four planner, and two apical, are negatively charged with two excess electrons (these are semi-covalent bonds, as discussed in the text).} and these split the five $d$ orbitals into two-sub-groups: $e_g$ and $t_{2g}$. Three orbitals ($xy,~yz,~zx$) form $t_{2g}$ set and these have lower energy as their lobs do not directly point towards negatively charged four planner oxygen atoms and two negatively charged apical oxygen atoms of the cage.  This leads to reduced Coulomb repulsion between electrons in these orbitals and six negatively charged oxygen atoms at the corners of the octahedron. Thus they will fill first,  accommodating six electrons in total. Now we have three electrons remaining out of nine. The two orbitals $3d_{x^2-y^2}$ and $3d_{3z^2-r^2}$ of the set $e_g$ can accommodate four electrons and  are both "un-happy" as their lobs point directly towards negatively charged oxygen atoms. Question is which one is more "un-happy" and which one is less "un-happy". Here come's Anderson's crucial insight. The most "un-happy" orbital is $3d_{x^-y^2}$ as its four lobs point directly towards negatively charged four planner oxygen atoms (figure 1) whereas $3d_{3z^2-r^2}$ face repulsion only from two negatively charged apical oxygen atoms along the z-direction. Thus two electrons will go into $3d_{3z^2-r^2}$ (lower energy orbital) and the last electron has to reside in $3d_{x^2-y^2}$ orbital (higher energy orbital). And it is {\it this} electron which is responsible for magnetism as it remains un-paired as well as for strange metal behavior on doping, and also for unconventional superconductivity. 

\begin{figure}[!h]
\begin{center}
\includegraphics[height=6cm]{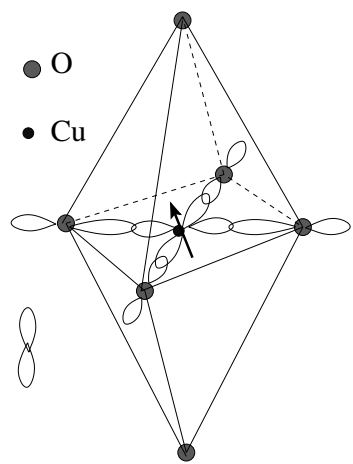}
\caption{The octahedron cage}
\end{center}
\end{figure}

This is the basic picture. There are other aspects. As four planner negatively charged oxygen atoms face lesser repulsion from one un-paired electron in $3d_{x^2-y^2}$ as compared to what two apical oxygen atoms face from two electrons in $3d_{3z^2-r^2}$, the four planer oxygens are pulled-in towards copper atom and apical oxygen atoms are pushed out. Thus the octahedron becomes distorted and  pointy. This is the Jahn-Teller (JT) distortion.  The presence of JT distortion is an unequivocal signature that un-paired electron resides in $3d_{x^2-y^2}$ orbital. 

Now the question is how the above picture is modified when hybridization is considered. Here it is important to understand the nature of bonding in Cuprates. There are two extreme kinds of chemical bonds: covalent bonds and ionic bonds.  In cuprates the bonding is intermediate i.e., semi-covalent or partly ionic.  Hybridization leads to bonding and anti-bonding orbitals. The highest hybrid orbital to be filled finally remains $Cu~3d_{x^2-y^2} -- O 2p_\sigma$ anti-bonding orbital. This picture is based on the fact that in Cuprates bonding has great strength (very high melting points) which is derived from semi-ionic character\cite{ander1}. 

What will happen when the system is hole doped i.e., when electrons are removed? The lone electron in $3d x^2 - y^2$ remains "un-happy" as it still has to face Coulomb repulsion from four negatively charged planner oxygen atoms. {\it Thus on hole doping, it is THIS electron which is removed. Therefore we only have one-component and one-band system.} The other topic of Mott insulating behaviour of a lattice of un-paired electrons localized in $3 d_{x^2-y^2}$ orbitals is well understood\cite{ander1,van}. Putting another electron into half-filled $3d_{x^2-y^2}$ orbital costs energy and the configurations of the type $Cu~ 3d^8 -- Cu ~3d^{10}$ are not energetically favored. We conclude based on the nature of chemical bonding in Cuprates that these are one-component or one-band systems as far as low energy physics is concerned.

\section{Two-component picture}

David Pines and collaborators very strongly argue that there are two sub-systems in cuprates:  one with localized spins and the other having itinerant character (figure 2). It is argued  that d-p hybridization is not compete and it leads to localized copper spins and mobile $p$ holes\cite{pines}. Hole doping does not remove the un-paired electrons from Cu $3d_{x^2-y^2}$ orbitals rather electrons are removed from oxygen $p$ orbitals. This leads to two types of electronic subsystems: one localized and the other mobile or itinerant. Further, NMR shift experimental data can be nicely explained using two-component model. Thus, in recent times this view seems to become dominant in the high-Tc community.\footnote{In the first decade after 1986 one-component view was more popular, but afterwards two-component view became more popular due to NMR shift experiments.}  Before we briefly review this analysis in the following paragraphs we first review the status of the literature in this field. 
\begin{figure}[!h]
\begin{center}
\includegraphics[height=2cm]{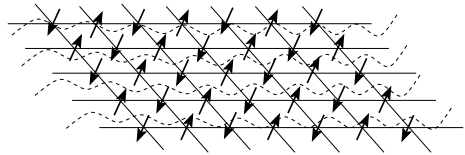}
\caption{The two-component view: Localized d-electrons (arrows) on Cu provide the pairing "glue" to itinerant or mobile p-band electrons from oxygen (wavy lines).}
\end{center}
\end{figure}

One component view was put forward in the RVB model by Anderson\cite{ander2}.  It was used early on in the investigations of Mila and Rice\cite{mila}, and Shastry\cite{shas}, and the famous Mila-Rice-Shastry Hamiltonian came into existence.  The celebrated Millis-Monien-Pines (MMP) phenomenological model for NMR relaxation in YBCO is motivated along these lines. The dichotomy between Korringa type behaviour of oxygen spin-lattice relaxation and non-Korringa type behaviour of the Cu spin-lattice relaxation is explained within one component  model using the idea of commensurate AFM spin correlations that leads to a certain kind of symmetry that causes the spin fluctuations at Cu site to vanish for oxygen relaxation\cite{mil}.

One of the first blow to the one-component view came from Walstedt, Shastry, and Cheong\cite{wal} in which they could not rationalized the dichotomy between Cu relaxation and O relaxation in LSCO within the one-component scenario of MMP.\footnote{However, refer to\cite{ul} for a different aspect related to coherence in hyperfine fields.} Recently, extensive investigations by Haase and collaborators strongly support the two-component view. Some selected references are\cite{haase,haase2,haase3}.

Next, we consider NMR shift experiments that points towards the two component view.

\section{Experimental support of the two-component picture}

NMR shift experiments are based on the "shielding" effect of un-paired electrons around a selected nucleus in which the nucleus will not "see" the externally applied magnetic field ($H_0$) rather it "sees" a modified field or an effective field $(H_{eff})$.  This changes the magnitude of the Zeeman splitting of the energy levels of the nucleus thus shifts the NMR resonance frequency. This can be accurately measured.\footnote{For other technical details of NMR shift experiments, reader is referred to dedicated literature\cite{haase,abra,sli,allo}.} This "shielding" due to un-paired electrons in technical literature is called Knight shift due to hyperfine interactions. For s-electrons it is Fermi contact type and due to p- or d-electrons, it is dipole-dipole type. Let us write the observed NMR resonance frequency as $\omega_{NMR}$. It can be written as
\beq
\frac{\om_{NMR}}{\gamma_n} = H_0 + \sum_k -A_{n,k}\la S_k\ra  + C_n.
 \eeq
Here, $H_0$ is the unscreened external magnetic field. $\la S_k\ra$ is the average value of the spin due to $k$th un-paired electron. $A_{n,k}$ is the hyperfine coefficient for $n$th nuclear site and $k$th electron. $C_n$ is the temperature independent part of the shift that comes from the orbital effects etc.

Let us assume, along with the supporters of the one component view, that there is one single temperature dependent susceptibility $\chi(T)$ which can be written as
\beq
M_{eff} = \mu_B g_k \la S_k\ra = \chi(T)H_0,
\eeq
where $M_{eff}$ is the effective magnetization. With this,  one can express the relative frequency shift in terms of a single temperature dependent susceptibility:
\beq
K_s = \frac{\om_{NMR} -\om_0}{\om_0} = - \sum_k \frac{A_{n,k}}{\mu_B g_k} \chi(T) + K_0.
\eeq
Here, $\om_0$ is the NMR resonance frequency without any shielding. Thus one concludes that $K_s \propto \chi(T)$. This means that shift must be a single function of temperature. {\it But this is not what is observed in general.} Let us take the case of $LSCO$ which we are considering.\footnote{Even YBCO under hydrostatic pressure show two temperature dependences\cite{haase}.}   An experimental data from one of the publications of Haase and collaborators is given in figure 3. 
\begin{figure}[!h]
\begin{center}
\includegraphics[height=5cm]{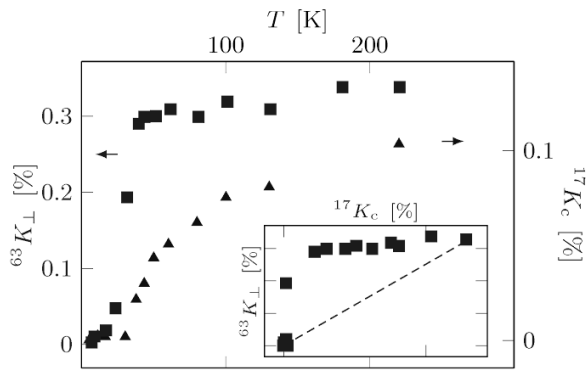}
\caption{Experimental data showing two different temperature dependences of shifts at Cu nucleus and O nucleus. This is against the one component view.  Figure courtesy\cite{haase}. }
\end{center}
\end{figure}
The shift at the Cu site is more or less temperature independent. It is suddenly reduced when $T_c$ is reached, whereas the shift at O is temperature dependent and smoothly decreases with decrease in temperature. Thus it is clear that it cannot be rationalized within a single temperature dependent susceptibility scenario. It can be understood as arising from two two fluids, one having more or less temperature independent susceptibility that is effective at Cu sites, and the other fluid having a temperature dependent susceptibility that is acting on O sites. This motivates the two component view (for more details refer to\cite{pines,haase}).

However, we argue differently. Next section presents our argument.

\section{An attempt to settle the debate}
A one-component system can lead to two-components: electrons in narrow bands with strong electron correlation  can show both the localized and the itinerant behavior at the same time. Thermal activation is the key.\footnote{This is not entirely new. Similar thermal activation ideas has been put forward in\cite{gor}.}  An electron temporally localized in the relevant  Cu d-orbital can convert to an itinerant electron through thermal excitation,  and reverse process also happens where an itinerant electron converts to a local one.  This thermal activation occurs over a "spin gap barrier" of magnetic origin. {\it The physical picture of our model consists of temporally localized states in Cu $d-$orbitals and a band of mobile states separated from the temporally localized states with a spin gap.} So, both types of the carriers can "emerge" in a single narrow band where localization tendencies are present due to Coulomb interactions,  and mobile states are separated from localized states via a "spin gap". This is some sort of {\it remanent of the zero doping Mott physics}  when the system is hole doped (figure 4). We mathematically model the situation in the following way.

\begin{figure}[!h]
\begin{center}
\includegraphics[height=2.5cm, width = 6.5cm]{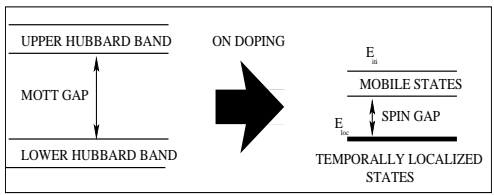}
\caption{The semiclassical model at finite doping is some sort of a remanent of the Mott physics at zero doping.}
\end{center}
\end{figure}
 
Let us consider $Cu-O$ lattice at zero doping. We have localized Cu spins (each Cu site with one un-paired electron). On hole doping let us say on $n_v$ fraction of sites electrons are removed. Now the system starts conducting.\footnote{Electronic conduction itself will have two channels: one through the tight band below the spin gap as the gap (as it turns out in Cuprates) has nodes in specific directions in momentum space. And the other channel is through excitation of carriers to upper band of mobile states above the spin gap. This excitation is thermal in nature. In our crude model we do not consider momentum dependence of the spin gap rather we use "an average spin gap". } Let, at any given instant of time, $n_{loc}$ be the fraction of localized electron in the lattice, and $n_{iti}$ be the fraction of itinerant electrons. These are time dependent quantities and keep on changing. But we have the obvious relation:
\beq
n_{loc} +n_{iti} + n_v =1.
\eeq 
The time evolution of the populations can be written as
\beq
\frac{d n_{loc}}{dt} = -P_{l\rta i} n_{loc} + P_{i\rta l} n_{iti} (1-n_{loc}).
\eeq
Here, $P_{l\rta i}$ represents the transition rate from local to itinerant behaviour, and $P_{i\rta l}$ is the reverse rate. The factor $(1-n_{loc})$ multiplying the last term represents the fact that an itinerant electron can become local only if vacant sites are available, whereas from going from local to itinerant no such constraint is there.  This is our method of imposing the Hubbard constraint in a soft way in this model. The two terms on the right hand side are noting but "loss and gain" of the local electron population.  Similarly
\beq
\frac{d n_{iti}}{dt} = - P_{i\rta l} n_{iti} (1-n_{loc}) + P_{l\rta i} n_{loc}.
\eeq
In the steady state
\beq
P_{l\rta i} n_{loc} = P_{i\rta l} n_{iti} (1-n_{loc}).
\eeq
Using the constraint (equation 4) and little algebra\footnote{Notice that if $n_v\rta 1$, that is all sites vacant (100 percent hole doping), $n_{loc}\rta 0$.}, we get
\beq
n_{loc} = \frac{1}{2}(\eta +2-n_v) - \sqrt{\frac{1}{4}(\eta +2 -n_v)^2 - (1-n_v)},
\eeq
and
\beq
n_{iti} = \sqrt{\frac{1}{4}(\eta +2 -n_v)^2 - (1-n_v)} - \frac{1}{2} (\eta + n_v).
\eeq
Here, $\eta$ is defined as
\beq
\eta = \frac{P_{l\rta i}}{P_{i\rta l}}.
\eeq

That is,  the local to itinerant transition rate divided by itinerant to local transition rate. To get the feel for numbers, let us assume that $n_v =0.1$ (that is 10 percent sites are vacant), let us take $\eta \simeq 1$.\footnote{In fact $\eta$ will be less than one, as the tendency of local to itinerant transition is suppressed as compared to the tendency of itinerant electron going local. This is due to short range antiferromagnetic correlations in localized electrons which tries to "hold them up" into the lattice (that is magnetic energy lowering while on localization). This energy is denoted by "an effective spin gap" in our model.} On plugging these numbers into equations (8) and (9) we get $n_{loc} \simeq 0.35$, and $n_{iti} \simeq 0.55$. In simple words, if we start with 100 sites with 100 Cu spins, then 10 percent vacant sites means we are left with 90 local spins and 10 vacant sites. If we switch on the time evolution, 55 electrons become itinerant and 35 remain localized. This equilibrium is not static, it is dynamic, localized and itinerant electrons keep on exchanging between themselves. {\it Thus we observe that a two component system does emerge from a one band and one  component system. }

To test the model we need to compute some observable from it and compare that with experiment. To this end, we will compute Knight shift from this model. But before we do that we need to compute the transition rates. This is done using the thermodynamical argument. Let $E_{loc}$ be the energy of temporally localized electrons and $E_{iti}(n_{iti})$ be the energy of the itinerant electrons which is a function of the itinerant electron number density ($n_{iti}$). Let $\Delta_{sg}$ be the spin gap. Populations obey the thermodynamic relations (refer to figure 4) 

\beqar
\frac{n_{loc}}{n_{loc} + n_{iti}} &=& e^{-\beta E_{loc}}.\nonumber\\
\frac{n_{iti}}{n_{loc} + n_{iti}} &=& e^{-\beta (E_{loc} + \Delta_{sg} + E_{iti}(n_{iti}) )}.
\eeqar

Their ratio gives

\beq
\frac{n_{loc}}{n_{iti}} = e^{\beta(\Delta_{sg} + E_{iti}(n_{iti}))}.
\eeq
For a 2D system $E_{iti}(n_{iti}) = \frac{\hbar^2}{4\pi m} n n_{iti} $. Here $n$ is the number of electrons per unit area and $n_{iti}$ is the fraction of itinerant electrons. If $a$ is the Cu-Cu bond length, then $n= 2/a^2$ with one $3d_{x^2-y^2}$ electron per Cu atom. Collecting all this, and writing $n_{loc}$ in terms of $n_{iti}$ using the constraint (equation 4 ) we get

\beq
ln\left(\frac{1-n_{iti} -n_v}{n_{iti}}\right) =\frac{1}{k_B T}\left(\Delta_{sg} + \frac{\hbar^2}{2 \pi m a^2} n_{iti}\right)
\eeq

This is one of our main result. The temperature dependence of $n_{iti}$ can be calculated from a numerical solution of the above implicit equation for $n_{iti}$ and it is presented in the Appendix figure (6). $n_{iti}$ increases with temperature as expected (thermal activation populates the itinerant band). The temperature dependence of the ratio of transition rates $\eta$ can be computed from equations (8) or (9). This is also given in the appendix figure (6). Here we deal with the important quantity which is the magnetic susceptibility of this semi-classical "one-level and one-band" model. For that we can deduce the temperature dependence of the Knight shift and can compare with  the experiment.

\begin{figure}[!h]
\begin{center}
\includegraphics[height=4cm]{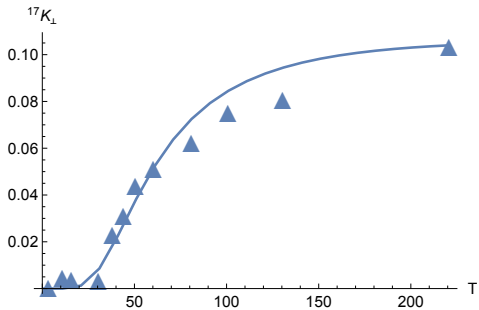}
\caption{Our theory (solid line) for $K^{17}_{\perp}(T)$ agrees reasonably well with the experimental data (triangle)\cite{haase}. The best fit value using least squares gives $\Delta_{sg} \simeq 9.3~eV$ which is in good agreement with its value found by  change in the slop of electrical resistivity and Nernst effect signal\cite{cyr}. There, the temperature $T^*$ deduced is about $125~K$ at $x=0.15$ which corresponds to $k_B T \simeq \Delta_{PG}$ about $10.8~meV$ for $\Delta_{PG}$ (figures 5 and 10 in\cite{cyr}). }
\end{center}
\end{figure}

Consider the magnetic susceptibility of itinerant electrons. In simple metals it is the Pauli susceptibility which is temperature independent. Temperature independence of the Pauli susceptibility in metals comes form the well known relation $\mu_B H << k_B T<< E_F$. In systems where Fermi energy is small and is comparable of $k_B T$, and/or  there is a gap in the excitation spectrum, the "Pauli susceptibility" will depend on temperature. In the next paragraph we compute the magnetic susceptibility due to the itinerant part in our "one-level and one-band" model,  in which number density ($n_{iti}(T)$),  and thus Fermi energy are temperature dependent quantities!

Magnetic susceptibility can be written as

\beqar
\chi_{iti}(T) &=& 2 \mu_B^2 ~~ \lim_{H \to0} \int_{\Delta_{sg}}^\infty d\ep g(\ep)\nonumber\\
&\times& \frac{f(\ep-\mu_B H) - f(\ep + \mu_B H)}{2 \mu_B H}.\nonumber\\
\eeqar

Where $g(\ep)$ is electron density of states ($\frac{4\pi m}{\hbar^2}$) in 2D. $f(\ep)$ is the Fermi function and $H$ is the external magnetic field. A straightforward calculation leads to
\beq
\chi_{iti}(T) = \frac{8\pi m \mu_B^2}{\hbar^2}\left(\frac{1}{e^{\beta(\Delta_{sg} -E_{iti}^F(T))} + 1}\right).
\eeq

Here $E_{iti}^F(T) =  \frac{\hbar^2}{2 \pi m a^2} n_{iti}(T)$. This is our main result. 

For the computation of Knight shift we assume that Knight shift at oxygen site is affected mainly by this component. There will be some effect on oxygen shift due to temporally localized spins on Cu atoms (transferred interactions). We assume that this effect is sub-dominant and oxygen shifts are mainly affected by the above computed susceptibility due to the itinerant part. Thus we set $K^{17}_{\perp}(T) \propto \chi_{iti}(T)$. Figure (5) shows the least square fitting of oxygen Knight shift computed from our model with that of experimental data of Haase et al\cite{haase}. The agreement is quite good. To compare the temperature evolutions of the Knight shift, the proportionality constant in $K^{17}_{\perp}(T) \propto \chi_{iti}(T)$ is normalized to the experimental data at the maximum temperature ($T\simeq 220~K$). We used $\Delta_{sg}$ as fitting parameter and best fit value gives $\Delta_{sg} \simeq 9.3~meV$. This is in reasonable agreement with that found through other methods\cite{cyr}.  The pseudogap  temperature $T^*$ deduced is about $125~K$ at $x=0.15$ (figures 5 and 10 in\cite{cyr}) which corresponds to $\Delta_{PG} \simeq 10.8~meV$ through $k_B T \simeq \Delta_{PG}$.

Theoretical modeling of the temperature dependence of the Cu Knight shift is beyond the scope of this investigation. The contribution in this case comes mainly from temporally localized electrons which have dynamically fluctuating antiferromagnetic correlations. As can be seen from figure (3) it is more or less temperature independent in the case of LSCO while this shift is temperature dependent in the case of YBCO at ambient pressure.\footnote{The shift when magnetic field is parallel to c-axis (that is $K^{63}_{//}$) shows large variations with temperature even within a given system. Thus it is less reliable than $K^{63}_{\perp}$ (magnetic field perpendicular to the c-axis or in the ab-plane).} Thus there is system to system variation even in the case of Cu ($K^{63}_{\perp}$) shift. Proper understanding of it requires details of the electronic structure variations from one system to another. This problem remains theoretically open\cite{haase}. 

However, we offer a qualitative understanding of it in the following way. There are two opposing tendencies acting at the Cu sites. The spin gap leads to lowering of Cu Knight shift on reducing temperature (as in figure 5). Whereas temporally localized spins of Cu sites leads to susceptibility that may scale like $\frac{1}{T}$ (Langevin-Curie type). This tends to increase with lowering temperature. Thus these two opposing tendencies may cancel each other and may lead to more or less temperature independent Knight shift at Cu site as seen in figure 3 (upper squares). It suddenly reduces when the system becomes superconducting. In this way a qualitative understanding can be obtained. However, a quantitative theory is much needed.

\section{Summary}

We conclude that un-paired electrons in $Cu~3d_{x^2-y^2}--O~2p$ hybrid orbitals and a narrow band formed by them are responsible for both the localized behaviour and the itinerant behaviour\cite{ander4}. An electron localized in that orbital can convert to an itinerant electron in a timescale of the order of $\frac{\hbar}{\Delta_{sg}} \sim 100~femto-sec$ (for $\Delta_{sg}\sim 10 ~meV$) and reverse process can also happen on similar timescales. {\it Thus a two-component system evolves from a manifestly  one-component system through thermal excitation.}  We put forward a semiclassical model to this end which takes into account this conversion process  through thermal activation and deactivation. Magnetic susceptibility and Knight shift is calculated in such a model. We find a reasonable agreement between theory and experiment for oxygen shifts data.

\section{Acknowledgment}
The author would like to thank B. Sriram Shastry for helpful comments.

\section{Appendix}
Thermal evolution of $n_{iti}(T)$ is plotted in figure (6) below. 
\begin{figure}[!h]
\begin{center}
\includegraphics[height=3.5cm, width= 6cm]{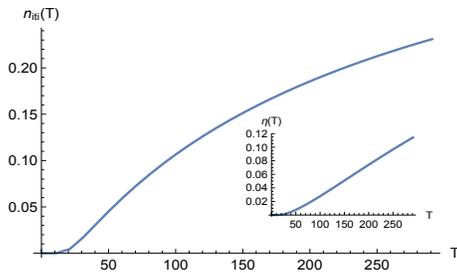}
\caption{Thermal evolution of $n_{iti}(T)$. It increases with increasing temperature as thermal excitation populates the itinerant band. Inset shows the temperature dependence of the ratio $\eta$ computed from equation (9) after solving equation (13).}
\end{center}
\end{figure}



\end{document}